\documentclass{article}

\usepackage{arxiv}

\usepackage[utf8]{inputenc} 
\usepackage[T1]{fontenc}    
\usepackage{hyperref}       
\usepackage{url}            
\usepackage{booktabs}       
\usepackage{amsfonts}       
\usepackage{nicefrac}       
\usepackage{microtype}      
\usepackage{lipsum}		
\usepackage{graphicx}
\usepackage{doi}
\usepackage{float}
\usepackage{array}
\usepackage{enumitem}
\usepackage[square,sort,comma,numbers]{natbib}

\title{Sustainable, Local Socio-Economic Development Through Astronomy}

\author{ \href{https://orcid.org/0000-0002-9745-0504}{\includegraphics[scale=0.06]{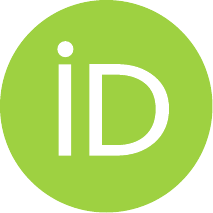}\hspace{1mm}Joyful E. Mdhluli}{ on behalf of the IAU Office of Astronomy for Development} \thanks{Visit our website, www.astro4dev.org or email: info@astro4dev.org} \\
	International Astronomical Union's Office of Astronomy for Development\\
	South African Astronomical Observatory\\
        Cape Town, South Africa\\
	\texttt{joy@astro4dev.org} \\
	\And
}

\renewcommand{\shorttitle}{\it{Socio-economic Development Through Astronomy}}

\hypersetup{
pdftitle={Sustainable, Local Socio-Economic Development Through Astronomy},
pdfsubject={astro-ph.IM},
pdfauthor={Joyful E.~Mdhluli},
pdfkeywords={Astronomy, Sustainable Development, Impact, Socio-economic Development},
}

\begin{document}
\maketitle

\begin{abstract}
Astronomy, often perceived as a distant or luxury science, holds immense potential as a driver for sustainable local socio-economic development. This paper explores how astronomy can create tangible benefits for communities through education, tourism, technology transfer, and capacity building. Using case studies from South Africa, Chile, Indonesia, and India, we demonstrate how astronomical facilities and initiatives have stimulated local economies, generated employment, supported small enterprises, and enhanced STEM participation, while simultaneously inspiring a sense of shared global heritage. The analysis identifies both successes and challenges, including unequal benefit distribution, limited local ownership, and sustainability gaps once external funding ends. Building on these lessons, we propose a practical framework/guidelines for designing, implementing, and evaluating astronomy-based community initiatives, rooted in participatory engagement and aligned with the UN Sustainable Development Goals (SDGs). This paper positions astronomy as a catalyst for inclusive growth, demonstrating that investment in the cosmos can translate into grounded, measurable benefits for people and places on Earth.
\end{abstract}


\section{Introduction}       
Astronomy is generally seen as a “luxury” science that only first world countries can afford. It is often believed that it is a science that requires expensive infrastructures and is of no use for economically disadvantaged countries, especially the many African countries that face socio-economic challenges. For this reason, astronomy very often appears as a low-priority science. However, the unique ability of astronomy to stimulate curiosity in the minds of young children and adults alike, as well as imagining possibilities, gives it a special place among the efforts to address development challenges.

In the hope of changing the current perspective and using astronomy as a tool to contribute to socioeconomic development in the world and especially in underdeveloped countries, the International Astronomical Union’s Office of Astronomy for Development (IAU-OAD) has funded several projects that aim to address some of the development challenges \cite{comment1}\cite{comment2}.

\begin{figure}[h!]
\centering
  \includegraphics[width=\textwidth]{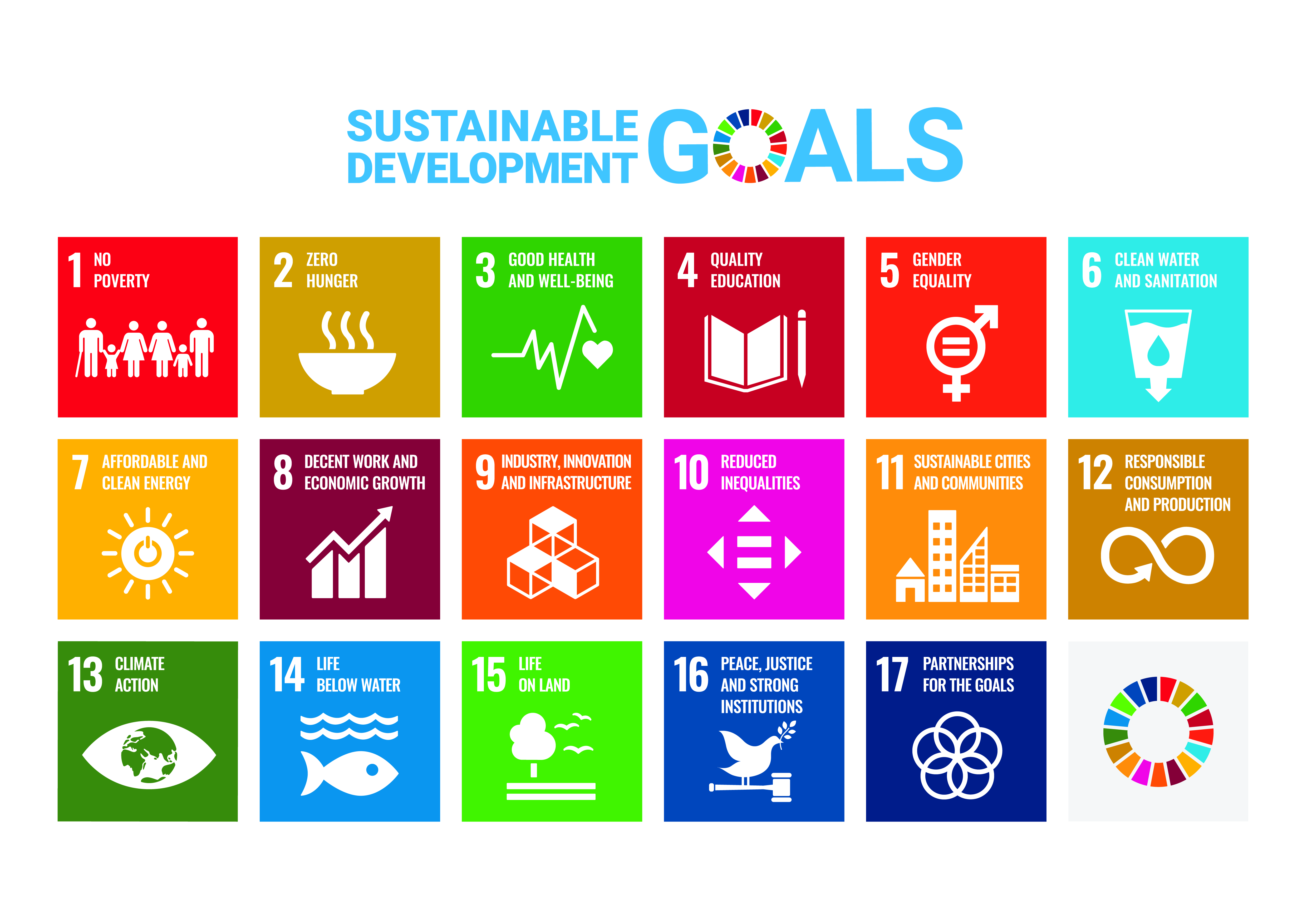}    
\caption{United Nations Sustainable Development Goals. \textit{Photo credit: UN Sustainable Development Goals} \cite{sdgs}}\label{un sdgs}
\end{figure}

In this paper, we provide a synthesis of some projects that aimed at using an astronomical facility as a hub within a small town or village to stimulate various associated socio-economic benefits for the local community, and provide a few guidelines for future such projects on how to optimise their outputs. We particularly focus on the specific case of ‘astrotourism' and the promotion of STEM education through astronomy, two types of projects that directly impact some of \href{https://www.globalgoals.org/}{United Nations Sustainable Development Goals} (UN SDGs) \cite{sdgs}, see Fig. \ref{un sdgs} for the SDGs. 

\section{About the International Astronomical Union's Office of Astronomy for Development}       
Astronomy serves as a powerful tool for addressing global challenges and promoting sustainable development in various domains, including scientific, technological, social, cultural, economic, and environmental aspects. Recognising the profound cultural and historical significance of astronomy, the International Astronomical Union's Office of Astronomy for Development (IAU-OAD) utilises its versatility to address diverse societal challenges. The OAD has established 11 Regional Offices (commonly referred to as ROADs) and Language Centres (commonly referred to as LOADs) around the world to further amplify the impact of astronomy for development. ROADs and LOADs share the OAD’s vision but tailor their activities to specific regional, cultural, or language needs.

The OAD’s efforts focus on several key areas where astronomy can contribute to positive change. These areas include enhancing educational outcomes, developing skills, promoting socio-economic development, and improving mental health and well-being. This focus translates into the development of partnerships with diverse organisations, governments, and communities to ensure equitable access to the benefits of astronomy. Ultimately, the OAD aims to leverage the resources and expertise of astronomy to contribute meaningfully to achieving the SDGs in socio-economic development, education, and well-being.

\begin{figure}[H]
    \centering
    \includegraphics[width=0.45\textwidth]{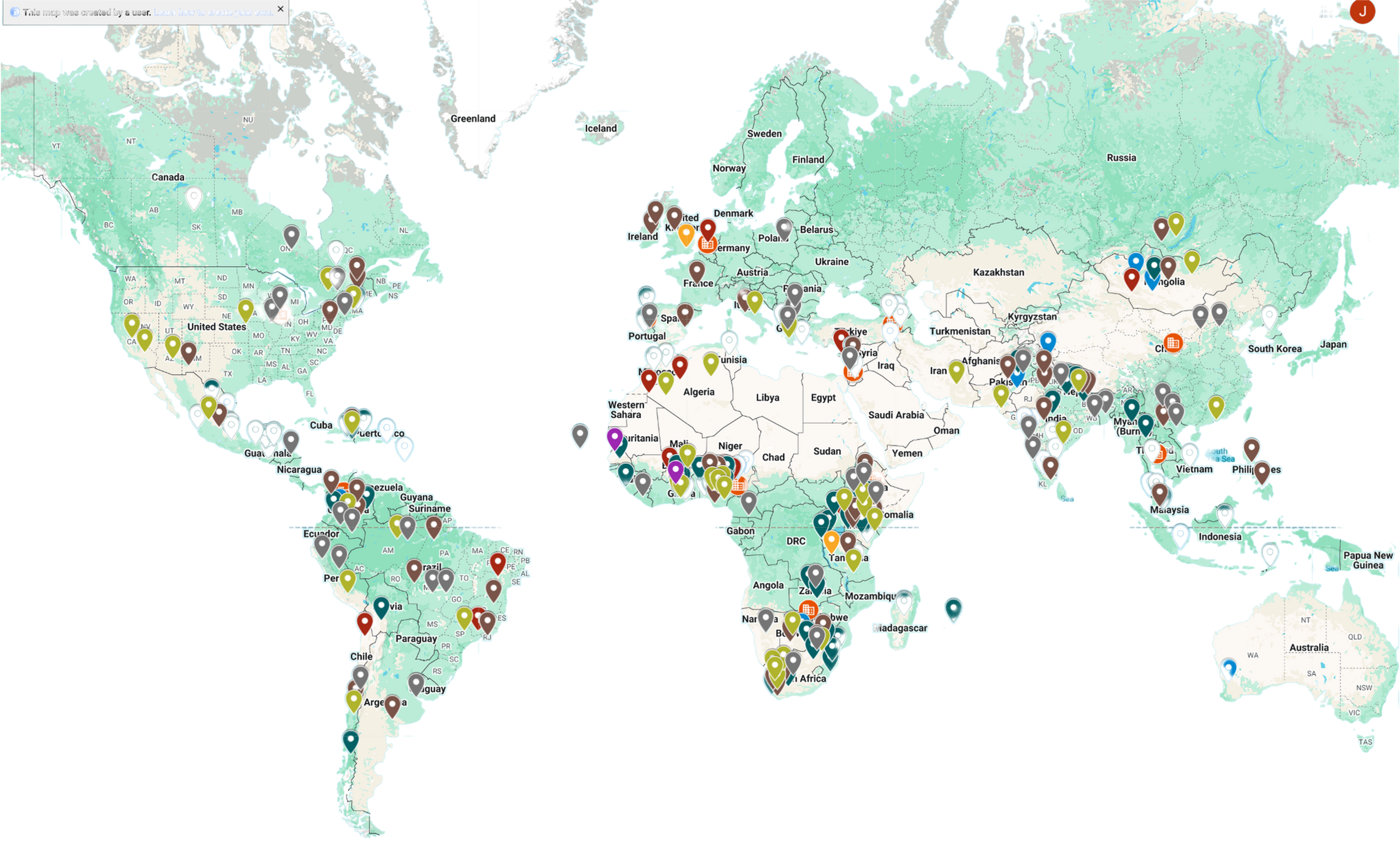}
    \includegraphics[width=0.45\textwidth]{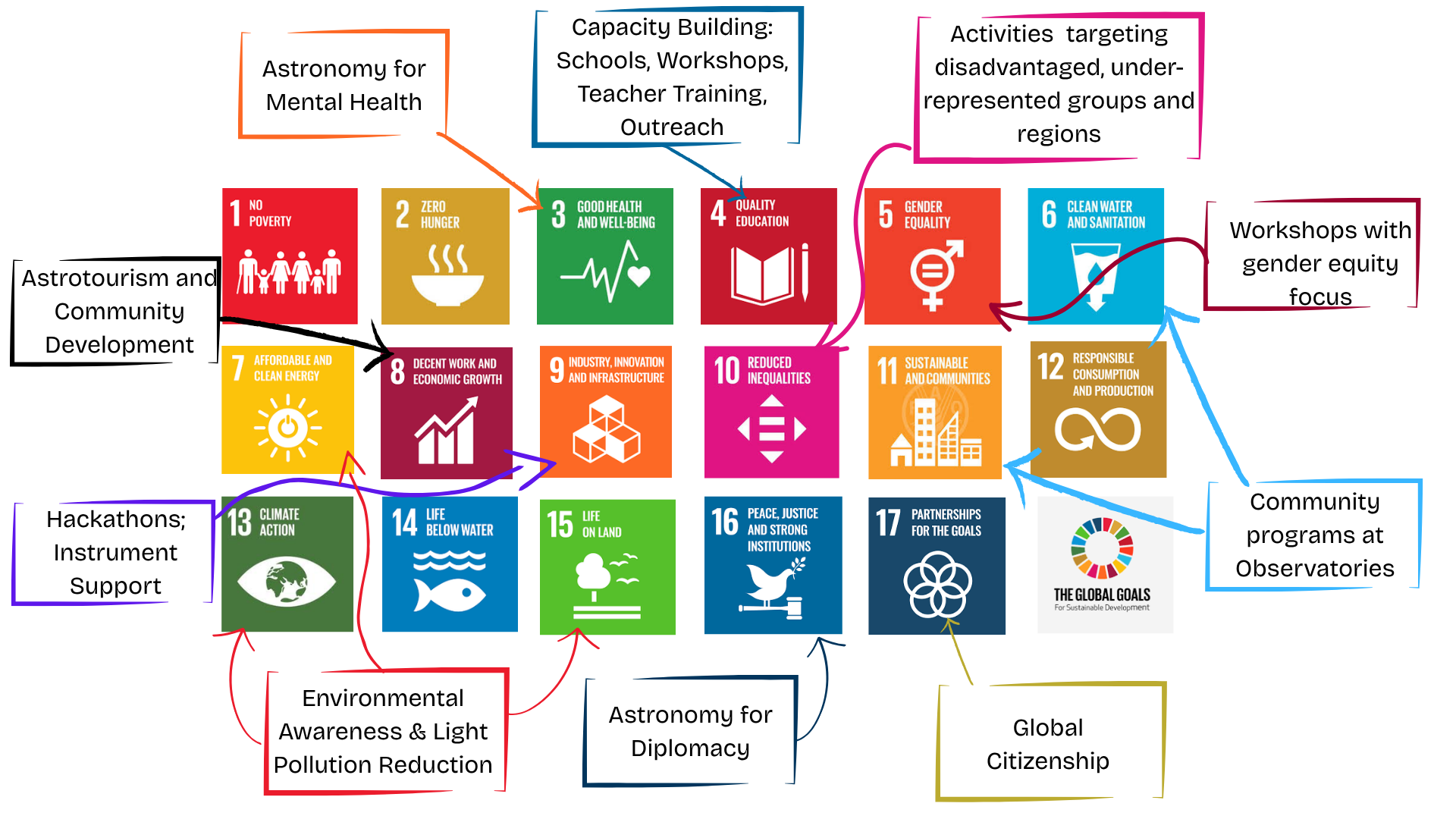}
    \caption{\textbf{Left:} Geographical distribution of ROADs, LOADs, and funded projects (2012–2024). \textbf{Right:} United Nations Sustainable Development Goals addressed by projects funded through the Call for Proposals.}\label{projects}
\end{figure}

The OAD achieves its objectives by strategically funding and coordinating projects that utilise astronomy to address challenges of sustainable development. Since 2012, the OAD has funded more than 200 projects in over 100 countries through its annual Call for Proposals.\footnote{https://www.astro4dev.org/cfp/} To ensure the success of funded projects, the project leaders are connected with the relevant ROADs or LOADs for additional support. As depicted in Figure \ref{projects} on the left, the OAD has a significant global presence.

Analysis of funded projects has revealed recurring themes, leading to the establishment of Thematic Areas. These thematic areas represent promising and successful astronomy-for-development concepts with the potential to scale up their impact globally. The thematic areas are:
\begin{itemize}
    \item Sustainable, local socio-economic development through Astronomy (e.g. astrotourism, observatories for communities, etc)
    \item Science diplomacy through Astronomy: Celebrating our Common Humanity (e.g. peace, post-conflict, partnerships, policy, etc)
    \item Knowledge and Skills Development (e.g. data science, data analysis, skills development, hackathons, etc)
    \item Addressing inequality (e.g. gender, geographic, ability, access, etc)
    \item Technology from Astronomy (e.g. software, water, solar, dark skies, etc)
\end{itemize}

Figure \ref{projects} on the right, highlights some of the SDGs addressed by projects funded through the annual Call for Proposals. This paper focuses on the first theme, “Sustainable, Local Socio-Economic Development Through Astronomy" \cite{comment1}\cite{comment2}.

\section{Case Studies from Past Projects}       
This section presents five past projects that were either funded, co-funded, or conducted in collaboration with the OAD. These case studies highlight the various lessons learned and demonstrate the socio-economic development impacts of these initiatives on targeted populations. Below, we summarise examples of local socio-economic impacts achieved through community projects conducted by organisations or observatories around the world.

\begin{figure}[H]
    \centering
    \includegraphics[width=0.58\textwidth]{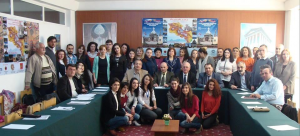}
    \includegraphics[width=0.35\textwidth]{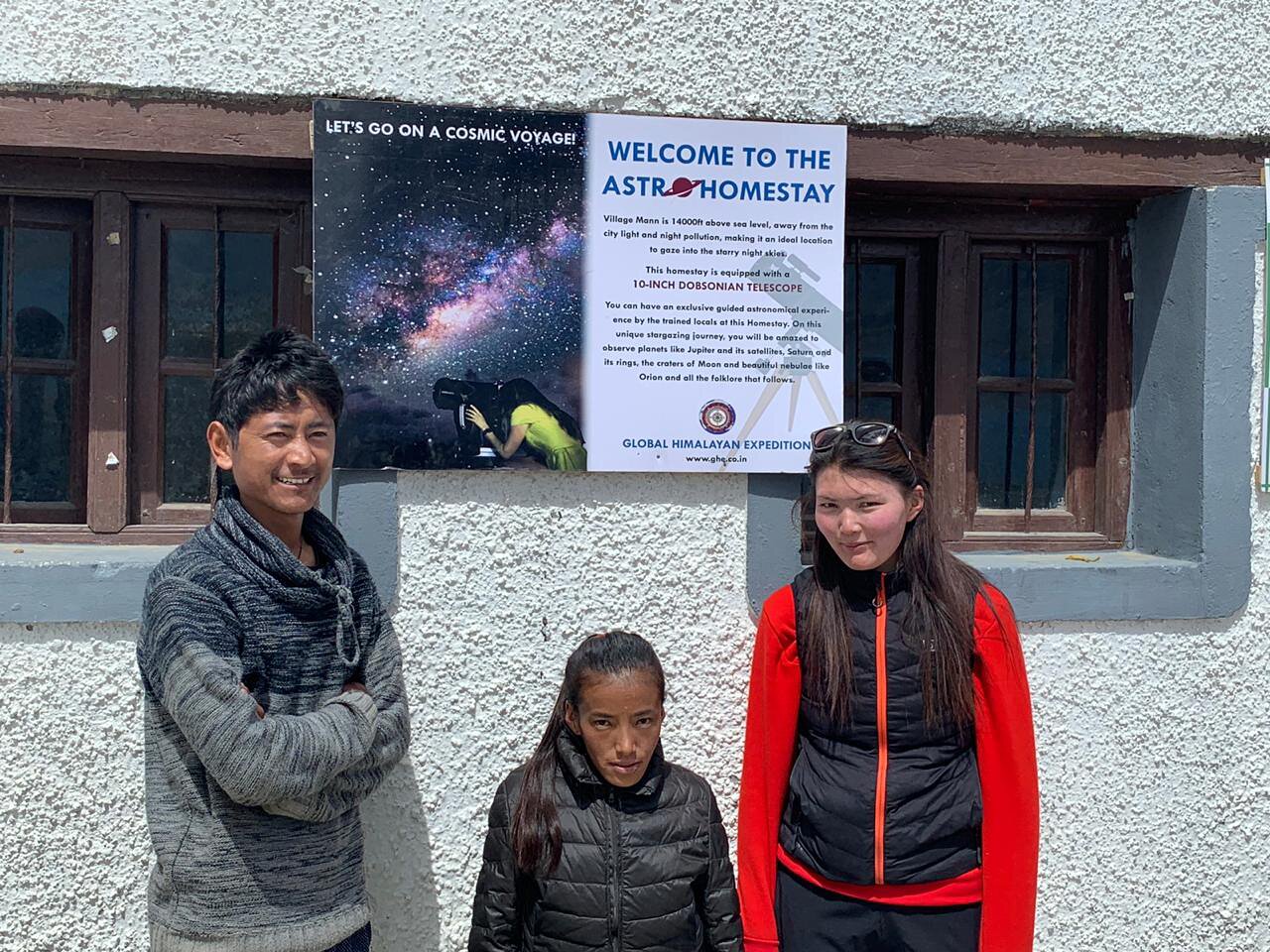}
    \caption{\textbf{Left:} Conference photo: “Scientific Tourism in the South West Asian Region” (March 2016, Armenian Institute of Tourism, Yerevan). \textit{Photo Credit: IAU SWA-ROAD.} \textbf{Right:} Villagers from Ladakh welcome tourists to their first AstroStay. \textit{Photo Credit: Astrostays/GHE.}}\label{examples}
\end{figure}

\begin{enumerate}
    \item \textbf{SALT Collateral Benefits Programme (SCBP), South Africa} \\
    The construction of the Southern African Large Telescope (SALT)\footnote{The Southern African Large Telescope is funded by a consortium of international partners from South Africa, the United States, Germany, Poland, India, the United Kingdom, and New Zealand.} in Sutherland led to the establishment of the SCBP (SALT Collateral Benefits Programme) to ensure the local communities benefit from the presence of the telescope. The programme focuses on education in science, technology, engineering, and mathematics (STEM), science communication and awareness, socio-economic development, and public engagement. Needs assessments for the programme were conducted through consultations with communities, stakeholders, and municipalities, addressing local challenges such as unemployment, alcoholism, drug abuse, and teenage pregnancy. 

    Initiatives under the programme include the creation of a Community Development Centre (CDC) in Sutherland and the enhancement of a visitor centre to boost tourism. Over four years (2015–2018), the observatory attracted more than 50,300 visitors to Sutherland, establishing SALT as \href{https://www.experiencenortherncape.com/visitor/experiences/saao-south-african-astronomical-observatory}{a major attraction} in the Northern Cape province. This activity has created local jobs and contributed to the socio-economic development of the area. A 2019 study revealed that, as of 2017, approximately 6\% of Sutherland's population was directly employed by the observatory, while 20\% reported deriving indirect benefits \cite{vorster}. The programme also emphasises STEM education through teacher training, school visits, learner workshops, and job-shadowing initiatives. In 2016 alone, over 46,000 school learners were reached through workshops, and more than 1,800 teachers participated in multi-day training workshops.
    \item \textbf{Community Development Around Timor Observatory, Indonesia}\\
    Partly funded by the IAU-OAD, the \href{https://astro4dev.org/category/tf3/community-development-around-timor-observatory/}{Community Development Around Timor Observatory} project partnered with the Indonesian Institute for Energy Economics (IIEE) to foster a symbiotic relationship between a proposed astronomical observatory in Timor, Indonesia, and the surrounding community. By 2016, a detailed survey had identified the needs of local populations near the observatory site. Engagements with local populations and authorities ensured the project aligned with community aspirations.
    The project implemented human capacity-building initiatives, including financial management training for public officers and STEM training for school teachers. The team emphasised the importance of addressing political and cultural nuances, building trust through public talks and participation in local festivities.
    \item \textbf{ALMA Observatory, Chile} \\
    The \href{https://www.almaobservatory.org/en/about-alma/alma-in-chile/benefits-for-the-community/}{Atacama Large Millimeter/submillimeter Array (ALMA) Observatory} has significantly contributed to Chile's development. It created local jobs (80\% of the staff are Chilean) and supported education through science and English programmes in a local school. These initiatives, implemented with support from specialists, improved students’ performance in national tests and gained local acclaim. ALMA also plans to build a visitor centre to promote outreach and enhance tourism. It collaborates with local museums and stakeholders to preserve cultural heritage, including an \href{https://www.almaobservatory.org/en/publications/the-universe-of-our-elders/}{ethno-astronomy project} designed in Spanish and English.
    \item \textbf{Astro-Tourism Development in South West Asia} \\
    The OAD-funded \href{http://iau-swa-road.aras.am/eng/AstroTourism/}{“Development of Astro-Tourism in South West Asia”} project, initiated in 2016, brought together organisations like the IAU South West and Central Asian ROAD and the Byurakan Astrophysical Observatory (BAO). Activities included visits to astronomical sites, conferences, and science camps. Outcomes included a promotional booklet on BAO and dedicated websites connecting tour agencies with astronomical sites. Figure \ref{examples} on the left, shows a photo from a related conference.
    \item \textbf{Astronomy for Himalayan Livelihood Creation, India}\\
    In collaboration with the IAU-OAD, the Global Himalayan Expedition (GHE) initiated an \hyperlink{https://astro4dev.org/blog/2017/12/19/overview-of-project-15/}{Astro-Tourism Program} to create sustainable livelihoods in unelectrified Himalayan villages. By setting up solar micro-grids and training women in tourism and hospitality, the project established 55 astro-homestays, generating $135,000$ in revenue. These homestays offer tourists unique stargazing opportunities and foster intercultural exchange. Figure \ref{examples} on the right, shows a proud community in Ladakh, India.
\end{enumerate}

\section{Lessons Learned from Past Projects}
This section highlights some of the lessons learned from the projects discussed in the previous section.

\begin{figure}[h!]
\centering
\includegraphics[width=0.8\textwidth]{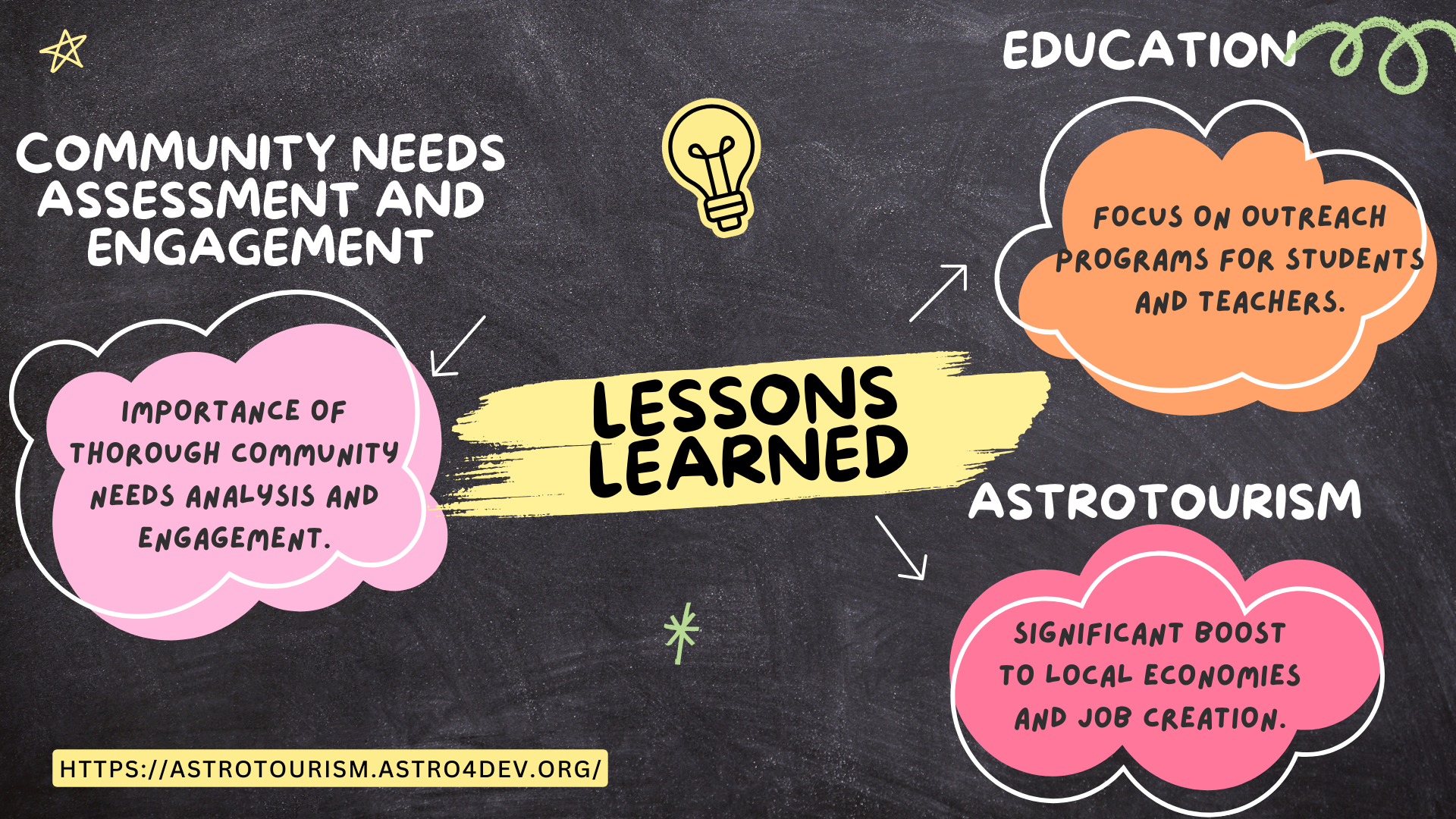}
\caption{Summary of Lessons Learned from Past Projects.}
\label{fig:lessons-learned}
\end{figure}

\begin{enumerate}
    \item \textbf{Community Needs Assessment and Engagement}\\
    The project focused on community development around the Timor Observatory emphasised the importance of thorough community needs analysis and active engagement. This approach allowed the project team to identify challenges faced by local communities and their expectations from the observatory. The needs analysis was conducted through an extensive study that included surveys, interviews with targeted communities, and demographic data collection from local or regional authorities. 

    This method proved effective despite challenges such as the time and human resources required. Engaging local populations is vital for building trust and efficiently identifying community needs, ensuring the success of development initiatives.

    \item \textbf{Astrotourism}\\
    The projects reviewed demonstrate the significant potential of astrotourism to boost tourism and socioeconomic development in rural areas. For example, in Sutherland, South Africa, the number of tourist accommodations has grown dramatically since the establishment of the observatory and the SCBP. This has generated substantial revenue and job creation, combating unemployment-related social issues such as alcoholism and drug abuse.

    However, challenges remain. Benefits of astrotourism often disproportionately favor wealthy minorities or non-locals with financial means to start businesses, as seen in Sutherland. Additionally, there is limited interaction between locals and tourists, which hinders cultural exchange and community involvement. Initiatives like the astro-homestays in the Himalayan project could address these issues by fostering direct community-tourist interaction and enabling disadvantaged locals to generate income.

    \item \textbf{Education}\\
    Educational initiatives associated with astronomical projects must prioritise schools in the surrounding communities. Outreach activities should target both learners and teachers. Teachers, in particular, need training in astronomy and related fields to inspire students and sustain interest in STEM disciplines. These efforts can build long-term interest in science and enhance the impact of educational programs in the community.
\end{enumerate}

\section{Guidelines for Future Projects}  
Developing astronomy-related projects with sustainable socio-economic benefits requires a strategic approach. These guidelines provide a roadmap to ensure that projects are impactful, equitable, and aligned with long-term development goals.

\begin{figure}[h!]
\centering
\includegraphics[width=0.85\textwidth]{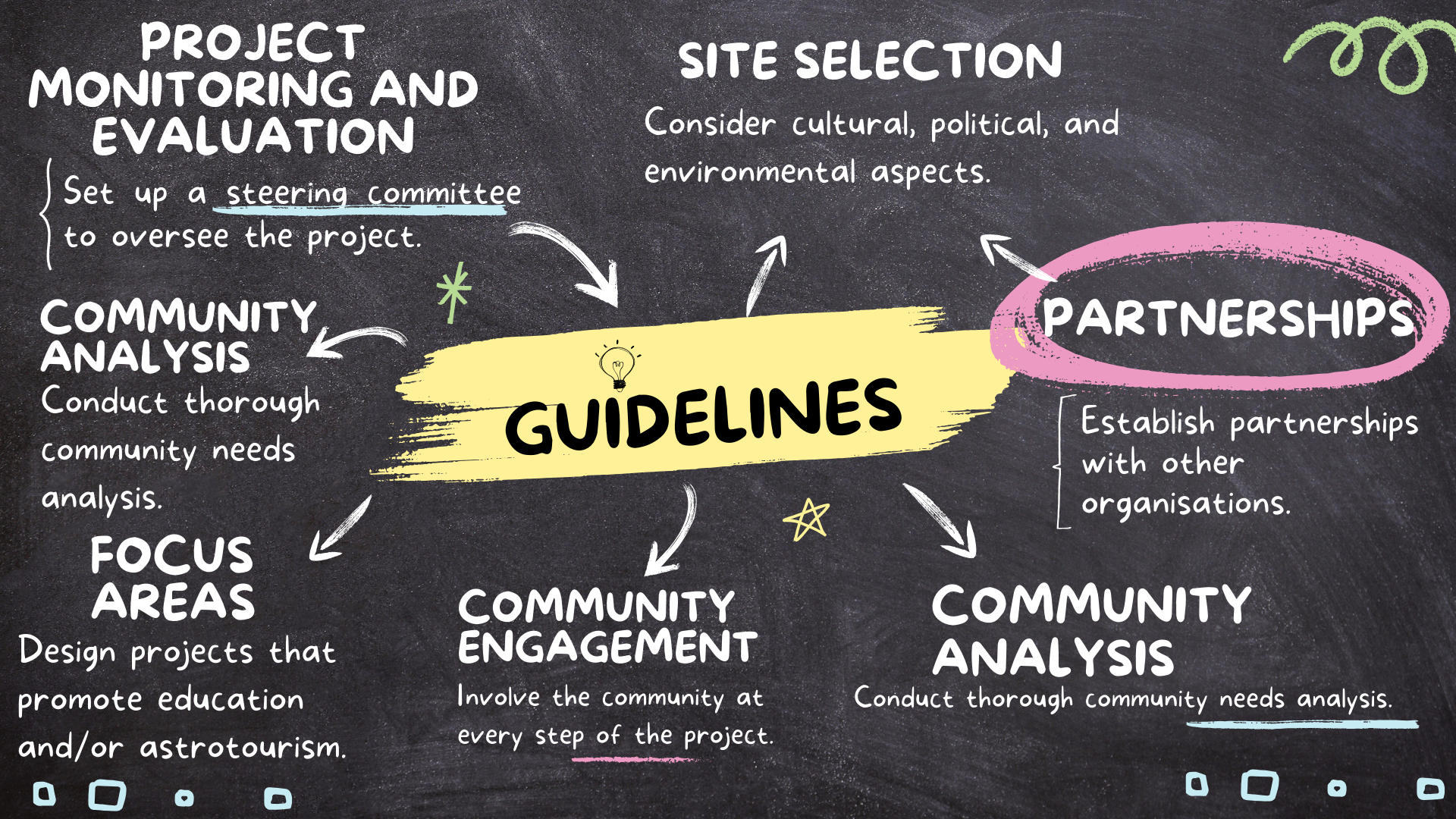}
\caption{Summary of Guidelines for Future Projects.}
\label{fig:guidelines-future-projects}
\end{figure}

\begin{enumerate}  
    \item \textbf{Site Selection}\\
    Selecting a site for an astronomical facility requires careful consideration of cultural, political, and environmental factors. The construction should not degrade the ecosystem or interfere with local cultural activities. Key steps include:
    \begin{itemize}
        \item Consulting local authorities and obtaining consent before finalising the site.
        \item Ensuring the site's safety and political stability, as conflicts can disrupt operations (e.g., a site in Burkina Faso had to be relocated due to terror attacks).
        \item Addressing technological constraints like light pollution for optical telescopes or radio interference for radio telescopes.
        \item Verifying accessibility with adequate road connections to nearby towns or villages.
    \end{itemize}

    \item \textbf{Community Analysis}\\
    Conducting a community needs analysis is essential to identify resources and tools that benefit local populations. This step should guide the size and nature of the facility and its integration into local development. Key questions include:
    \begin{itemize}
        \item What facilities exist in the area, and how can they support the astronomical project?
        \item Are there schools nearby? What are the literacy rates and age demographics of the population?
        \item Can the facility attract visitors and create jobs for the local community?
    \end{itemize}
    Guidelines for needs assessment include:
    \begin{itemize}
        \item \textbf{PESTLE Analysis:} Evaluate political, economic, sociological, technological, legal, and environmental factors.
        \item \textbf{SWOT Analysis:} Identify strengths, weaknesses, opportunities, and threats, integrating them into the PESTLE framework.
        \item \textbf{Surveys:} Conduct formal or informal surveys to gather local insights and estimate willingness to pay (WTP) for services.
    \end{itemize}

    \item \textbf{Community Engagement}\\
    Engaging local communities throughout the project lifecycle - design, implementation, and monitoring - is critical. Regular meetings, representation in project teams, and forums for community input help build trust and ensure alignment with local needs.

    \item \textbf{Feasibility Analysis}\\
    Assessing the practicality and expected returns of the project ensures its viability and sustainability. Key considerations:
    \begin{enumerate}
        \item \textbf{Project Returns:} Projects with financial or measurable non-financial returns (e.g., reduced youth drug abuse) are more likely to be sustainable.
        \item \textbf{Demand Analysis:} Use survey data and other indicators (e.g., HDI, education levels) to model demand and inform decision-making.
    \end{enumerate}

\item \textbf{Partnership}\\
Besides the OAD, the project team must seek to establish partnerships with other related organisations at regional, national, or international levels. The goal is to enable collaborations with organisations specialising in fields such as astronomy, tourism, community engagement, or evaluation, and to learn from their experiences. It is also important to partner with government or private agencies that can assist in providing essential resources such as electricity and water. Examples of potential partners include other astronomical facilities (such as observatories), research institutions, tourism and travel agencies, etc. Furthermore, if the project includes an education component, forming partnerships with local schools or higher education institutes is strongly encouraged.

\item \textbf{Project Monitoring \& Evaluation}\\
To ensure the sustainability of the project, consistent monitoring and evaluation are essential. A steering committee can be set up to supervise project activities and manage overall operations. This is particularly important as it evaluates the project's impact on the targeted local communities. More information on this step can be found on \cite{monitoring} and on the \href{https://astro4dev.org/monitoring-evaluation/}{OAD webpage}.

\item \textbf{Possible Objectives of Community Projects}\\
There are several ways to design projects that benefit communities. For example, these can focus on promoting education and/or astrotourism. Below, we provide additional, more technical guidelines specific to some of these focuses.
\begin{enumerate}
\item \textbf{Education}\\
The first step in designing a development program to promote STEM education using an astronomical facility is deciding whether the program will primarily target public outreach for the general public and school learners or focus on higher education through the development of an astronomy curriculum and research projects.

\begin{itemize}
    
\item \textbf{Public Outreach}\\
If the focus of the project is to promote STEM-related fields through public outreach activities, then the site designated to host the astronomical facility must be accessible to the general public. The facility should serve as a rallying point for STEM fields in the targeted community. Particular attention should be given to engaging young girls and underrepresented minorities in STEM fields.

\item \textbf{Tertiary Education Projects}\\
The first step in this category is to clearly define the scientific objectives and types of projects that will be conducted using the telescope. For this, collaboration with a (preferably local) university or tertiary education institute is essential, as such projects are intended for university students. This category also requires careful site selection, particularly regarding sky quality, as good science cannot be conducted if light pollution remains an issue. Activities developed under this category should focus on student training through research projects. Experience with the \href{http://www.ast.uct.ac.za/ast/teaching-telescopes/tony-fairall-teaching-observatory}{Tony Fairall teaching telescope} at the University of Cape Town demonstrates that a telescope with a diameter of 10-14 inches, equipped with a CCD camera, can be effectively used.
\end{itemize}

\item \textbf{Astrotourism}\\
A simple Google search with “astrotourism” as the keyword yields about 49,000 results, featuring several articles and websites promoting tourist destinations. There are even dedicated websites (e.g., \href{https://www.astrotourism.com/}{astrotourism.com}, \href{https://chile.travel/en/what-to-do/astrotourism}{chile.travel}, \href{https://www.andalucia.org/en/astrotourism-star-tourism/}{andalucia.org}) that provide information about astronomical facilities and stargazing sites, akin to hotel booking websites. This sector has gained significant popularity worldwide in recent years, drawing increasing numbers of tourists globally. For example, a New York Times \href{https://www.nytimes.com/2018/09/03/travel/stargazing-trips.html}{article} published in September titled “Your Next Trip? It’s Written in the Stars” explores the field's growing popularity, particularly across the United States, Canada, and Mexico. The article acknowledges the efforts of organisations such as the \href{https://darksky.org/}{DarkSky International} and the \href{https://www.rasc.ca/}{Royal Astronomical Society of Canada (RASC)} to preserve the night sky by designating dark sky parks and reserves across North America and Europe. It also highlights astrotourism sites, parks, resorts, and public festivals in the Americas, demonstrating the growing influence of the astrotourism business on the future of tourism. Organisations such as the \href{https://en.fundacionstarlight.org/}{Starlight Foundation} work to protect the night sky, promote astronomy's cultural diffusion, and foster local sustainable economic development through astrotourism (\href{https://www.fundacionstarlight.org/docs/files/78_declaracion-sobre-la-defensa-del-cielo-nocturno-y-el-dereho-a-la-luz-de-las-estrellas-ingles.pdf}{Declaration of La Palma or Starlight Declaration}).

Astrotourism, defined as a form of tourism that utilises unpolluted night skies and scientific knowledge for astronomical, cultural, or environmental activities \cite{fayos1}, is well-suited for community development. Unlike many other forms of tourism, it emphasises the conservation of natural resources, aligning with sustainable development goals.

To ensure an astrotourism program thrives, a professional approach to both the destination site and product management is essential \cite{fayos}. If this is the main theme of the astronomical facility, efforts must be made to ensure its sustainability. Like any tourism destination, a three-step analysis and policy process is necessary for a successful astrotourism program \cite{fayos}\cite{unwto}:

\begin{itemize}
\item A green paper of the site presenting a detailed inventory of available resources. This phase identifies existing products and services beneficial to the program's feasibility and sustainability. For instance, ensuring the quality of night skies from the observatory and sufficient scientific knowledge or human resources to support the project. It is also crucial to engage local authorities, community leaders, and potential stakeholders during this phase.

\item A white paper for strategic decision-making, selecting necessary resources, support services, and tourism products. This phase includes a comprehensive analysis of the astrotourism program's sustainability and robustness. The PESTLE and SWOT analyses described above should be conducted at this stage, with a focus on socioeconomic benefits for local communities.

\item A tourism policy plan, consisting of structured sub-programs to analyze market conditions, attract visitors, and meet the needs of both local communities and visitors. These sub-programs include:
\begin{itemize}
\item \textbf{Activities:} Design workshops and activities to attract and engage visitors.
\item \textbf{Data:} Continuously collect and analyze data to guide the program's sustainability.
\item \textbf{Sustainability:} Ensure the preservation of natural and cultural resources.
\item \textbf{Innovation:} Introduce new ideas to keep the program dynamic.
\item \textbf{Cooperation:} Collaborate with local and international organizations.
\item \textbf{Fundraising:} Explore funding sources for long-term facility maintenance.
\item \textbf{Knowledge and Human Capital:} Train tour guides and develop scientific content.
\item \textbf{Products and Promotion:} Promote the facility and associated activities to attract visitors.
\end{itemize}
\end{itemize}
The OAD has has developed a suite of openly accessible resources to support individuals and institutions interested in implementing astrotourism initiatives globally. These resources also encourage individuals and existing businesses to broaden their offerings to include activities that use the night sky as a backdrop, such as food experiences, wellness practices, and cultural exploration. An overview of the resources is available in reference \cite{resources} and the resources are available here: \cite{resource1}-\cite{resource3}
\end{enumerate}
\end{enumerate}
However, economic development in small towns may have negative effects, such as increased cost of living, worker exploitation, or light pollution. Although this document does not provide mitigation strategies, it is important to raise awareness of these issues and work with local authorities to address them.

\section{Conclusion}
The evidence presented affirms that astronomy can meaningfully contribute to sustainable socio-economic development when projects are community-driven, context-sensitive, and strategically aligned with local needs. From the visitor economy surrounding observatories to the empowerment of small-scale entrepreneurs through astrotourism, astronomy’s reach extends far beyond research and education, it can strengthen livelihoods, foster innovation, and enhance social cohesion. However, achieving this potential requires careful planning, inclusive governance, and long-term commitment to capacity building. Projects must go beyond infrastructure to cultivate local ownership, ensure equitable distribution of benefits, and establish mechanisms for continuous monitoring and evaluation. As demonstrated by this paper, astronomy offers more than knowledge of the universe: it provides a framework for imagining and enacting sustainable futures on Earth. Continued collaboration between scientists, policymakers, and communities will be essential to ensure that the light from the stars truly illuminates paths toward shared human development.


\begin{thebibliography}{9}
\bibitem{comment1}
McBride, V., Venugopal, R., Hoosain, M. et al. "The potential of astronomy for socioeconomic development in Africa." \textit{Nat Astron} 2, 511–514 (2018). https://doi.org/10.1038/s41550-018-0524-y

\bibitem{comment2}
Mdhluli, J.E., Takalana, C., Venugopal, R. et al. "Astronomy as a strategic driver for sustainable development." \textit{Nat Astron} (2025). https://doi.org/10.1038/s41550-025-02602-x

\bibitem{sdgs}
United Nations Sustainable Development Goals.\\ \href{https://sdgs.un.org/goals}{https://sdgs.un.org/goals}

\bibitem{vorster}
Vorster, J., \& Eigelaar-Meets, I. "Sutherland: Socio-economic characteristics." \textit{Cosmopolitan Karoo Research Report} (2019). https://cosmopolitankaroo.co.za/research-outputs-2/research-reports/

\bibitem{monitoring}
Mdhluli, J. E. \&  IAU Office of Astronomy for Development (2025). "Monitoring and Evaluating Astronomy-for-Development Initiatives." arXiv. https://doi.org/10.48550/arXiv.2512.18015

\bibitem{fayos1}
Fayos-Sol\`{a}, E., Marin, MC \& Jafari, J. "Astrotourism: No Requiem for Meaningful Travel." \textit{PASOS Revista de Turismo y Patrimonio Cultural} (2014). https://doi.org/10.25145/j.pasos.2014.12.048 

\bibitem{fayos}
Fayos-Sol\`{a}, E., Alvarez, M. \& Cooper, C. "Tourism as an instrument for development: A theoretical and practical study." \textit{ISBN 9780857246790} (2014). 

\bibitem{unwto}
UN Tourism Elibrary. "UNWTO Tourism Highlights, 2010 Edition." (2010). https://www.e-unwto.org/doi/pdf/10.18111/9789284413720 

\bibitem{resources}
Mdhluli, J. E. \&  IAU Office of Astronomy for Development (2025). "Astrotourism for Development: An Overview of Resources from the IAU Office of Astronomy for Development." arXiv. https://doi.org/10.48550/arXiv.2507.15827

\bibitem{resource1}
IAU Office of Astronomy for Development, Manikumar, S., \& Mdhluli, J. E. (2025). Night Sky Tourism: A Quick Start Guide. Zenodo. https://doi.org/10.5281/zenodo.15870920

\bibitem{resource2}
IAU Office of Astronomy for Development, Manikumar, S., \& Mdhluli, J. E. (2025). A Night Sky Tourism Guide For Existing Businesses. Zenodo. https://doi.org/10.5281/zenodo.15871078

\bibitem{resource3}
IAU Office of Astronomy for Development, Manikumar, S., \& Mdhluli, J. E. (2025). Night Sky Tourism for Communities Around Observatories. Zenodo. https://doi.org/10.5281/zenodo.15871175

\end{thebibliography}
\end{document}